\ifx\pdfoutput\undefined 
\documentclass{elsart}
\usepackage{amssymb}
\usepackage{graphicx}
\usepackage{epsfig}
\else
\documentclass[pdftex,fleqn]{elsart}
\usepackage{amssymb}
\usepackage[pdftex]{graphicx}
\DeclareGraphicsExtensions{.pdf,.png,.jpg,.mps}
\RequirePackage[colorlinks,hyperindex]{hyperref}
\usepackage{thumbpdf}
\fi

\newcommand{\lsim}{\mathrel{\mathop{\kern 0pt \rlap
  {\raise.2ex\hbox{$<$}}}
  \lower.9ex\hbox{\kern-.190em $\sim$}}}
\newcommand{\gsim}{\mathrel{\mathop{\kern 0pt \rlap
  {\raise.2ex\hbox{$>$}}}
  \lower.9ex\hbox{\kern-.190em $\sim$}}}
\newcommand{\gagamma}{g_{a\gamma\gamma}}

\begin{document}
\begin{frontmatter}

\title{Physics Potential and Prospects for the CUORICINO and CUORE Experiments}%

\author[label1]{C. Arnaboldi}, \author[label2]{F. T. Avignone III},
\author[label3,label4]{J. Beeman}, \author[label5]{M. Barucci},
\author[label6]{M. Balata}, \author[label1]{C. Brofferio},
\author[label6]{C. Bucci}, \author[label7]{S. Cebrian},
\author[label2]{R. J. Creswick}, \author[label1]{S. Capelli},
\author[label1]{L. Carbone}, \author[label1]{O. Cremonesi},
\author[label8]{A. de Ward}, \author[label1]{E. Fiorini},
\author[label2]{H. A. Farach}, \author[label8]{G. Frossati},
\author[label9]{A. Giuliani},
\author[label1]{P. Gorla},\author[label3,label4]{E. E. Haller},
\author[label7]{I. G. Irastorza}, \author[label3]{R. J. McDonald},
\author[label7]{A. Morales}\thanks{corresponding author: amorales@posta.unizar.es},
\author[label3]{E. B. Norman},
\author[label1]{A. Nucciotti},
\author[label9]{M. Pedretti}, \author[label7]{C. Pobes},
\author[label10]{V. Palmieri}, \author[label1]{M. Pavan},
\author[label1]{G. Pessina}, \author[label1]{S. Pirro},
\author[label1]{E. Previtali}, \author[label2]{C. Rosenfeld},
\author[label7]{S. Scopel},\author[label3]{A. R. Smith},
\author[label1]{M. Sisti},\author[label5]{G. Ventura},
\author[label1]{M. Vanzini}

\address[label1]{Dipartimento di Fisica dell'Universit\`{a} di
Milano-Bicocca e Sezione di Milano dell'INFN, Milan I-20136,
Italy}

\address[label2]{University of South Carolina, Dept.of Physics
and Astronomy, Columbia, South Carolina, USA 29208 }

\address[label3]{Lawrence Berkeley National Laboratory, Berkeley,
California, 94720, USA}

\address[label4]{Dept. of Materials Science and Mineral
Engineering, University of California, Berkeley, California
94720, USA }

\address[label5]{Dipartimento di Fisica dell'Universit\`{a} di
Firenze e Sezione di Firenze dell'INFN, Firenze I-50125, Italy }

\address[label6]{Laboratori Nazionali del Gran Sasso, I-67010,
Assergi (L'Aquila), Italy}

\address[label7]{Laboratorio de Fisica Nuclear y Altas Energias,
Universidad de Zaragoza, 50009 Zaragoza, Spain }

\address[label8]{Kamerling Onnes Laboratory, Leiden University,
2300 RAQ, Leiden, The Netherlands }

\address[label9]{Dipartimento di Scienze Chimiche, Fisiche e
Matematiche dell'Universit\`{a} dell'Insubria e Sezione di Milano
dell'INFN, Como I-22100, Italy }

\address[label10]{Laboratori Nazionali di Legnaro, Via Romea 4,
I-35020 Legnaro ( Padova ), Italy }

\centerline{(The CUORE COLLABORATION)}

\begin{abstract}
The CUORE (\emph{Cryogenic Underground Observatory for Rare Events}) experiment
projects to construct and operate an array of 1000 cryogenic thermal detectors
of TeO$_{2}$, of a mass of 760 g each, to investigate rare events physics, in
particular, double beta decay and non baryonic particle dark matter. A first
step towards CUORE is CUORICINO, an array of 62 of bolometers,
currently being installed in the Gran Sasso Laboratory. In this paper we report
the physics potential of both stages of the experiment regarding neutrinoless
double beta decay of $^{130}$Te, WIMP searches and solar axions.

\end{abstract}

\end{frontmatter}

\section{Introduction: The CUORE project}

Rare event Physics at the low energy frontier is playing a
significant role in Particle Physics and Cosmology. Examples of
such rare phenomena could be the detection of non-baryonic
particle dark matter (axions or WIMPs), supposedly filling a
substantial part of the galactic haloes, or neutrinoless double
beta decay. These rare signals, if detected, would be important
evidences of a new physics beyond the Standard Model of Particle
Physics, and would have far-reaching consequences in Cosmology.
The experimental achievements accomplished during the last decade
in the field of ultra-low background detectors have led to
sensitivities capable of searching for such rare events. Dark
matter detection experiments have largely benefited from the
techniques developed for double beta decay searches. Both types of
investigation, which have very close relation from an experimental
point of view, are the two main scientific objectives of CUORICINO
and CUORE.

Due to the low probability of both types of events, the essential
requirement of these experiments is to achieve an extremely low
radioactive background. For that purpose the use of radiopure
detector components and shieldings, the instrumentation of
mechanisms of background identification, the operation in an
ultra-low background environment, in summary, the use of the
state-of-the-art of low background techniques is mandatory. In
some of these phenomena, like the case of the interaction of a
particle dark matter with ordinary matter, a very small amount of
energy is deposited, and the sensitivity needed to detect events
within such a range of energies, relies on how low the energy
threshold of detection is. In addition, to increase the chances of
observing such rare events a large amount of detector mass is in
general advisable, and so most of the experiments devoted to this
type of searches are planned to have large detector mass, while
keeping the other experimental parameters (background, energy
thresholds and resolutions) optimized. On the other hand, the
recent development of cryogenic particle detection \cite{Mosca}
has led to the extended use of thermal detectors \cite{Fio84} to
take advantage of the low energy threshold and good energy
resolution that are theoretically expected for the thermal
signals. This detection technology has also the advantage of
enlarging the choice of materials which can be used, either as DM
targets or $2\beta$ decay emitters. After a long period of R\&D to
master the techniques used in cryogenic particle detectors, low
temperature devices of various types are now applied to the
detection of double beta decay or particle dark matter
\cite{Mor99}. A major exponent of this development is the MiDBD
(\emph{Milano Double Beta}) experiment
\cite{Pirro:2000fi,MiDBDfinale}, which successfully operated a 20
TeO$_2$ crystal array of thermal detectors of a total mass of 6.8
kg, the largest cryogenic mass operated up to now. With the
objective of going to larger detector masses and of improving the
sensitivity achieved in the smaller arrays, the CUORE
(\emph{Cryogenic Underground Observatory for Rare Events}) project
\cite{Fiorini:1998gj} was born some years ago as a substantial
extension of MiDBD. The objective of CUORE is to construct an
array of 1000 bolometric detectors with cubic crystal absorbers of
tellurite of 5 cm side and of about 760 g of mass each. The
crystals will be arranged in a cubic compact structure and the
experiment will be installed in the Gran Sasso Underground
Laboratory. The material to be employed first is TeO$_2$, because
one of the main goals of CUORE is to investigate the double beta
decay of $^{130}$Te, although other absorbers could also be used
to selectively study several types of rare event phenomenology.
Apart from the MiDBD experiment, already completed, a wide R\&D
program is under way in the framework of CUORICINO, a smaller and
intermediate stage of CUORE, which consists of 44 of the above
crystals (790~g in this case) and 18 smaller crystals (330~g) for
a  total mass of $\sim\,$40~kg. CUORICINO has been recently
installed in the Gran Sasso Laboratory (LNGS) and is, by far, the
largest cryogenic detector on the stage. The preliminary running
tests are encouraging \cite{Giulani}.

In the present work the prospects are presented for the CUORE and
CUORICINO experiments with respect to their double beta decay
discovery potential and their detection capability of WIMP and
axions. In section \ref{bc} considerations of the expected
background levels in the energy regions of interest are made
taking into account all relevant sources. The theoretical
motivation of neutrinoless double beta decay experiments as well
as the experimental prospects of both CUORICINO and CUORE are
shown in section \ref{dbd}. In the same way, the physics potential
of the experiments is discussed for direct detection of WIMPs in
section \ref{wimps} and for solar axion searches in section
\ref{axions}.

\section{Background considerations} \label{bc}

From the radioactive background point of view, thermal detectors
are very different than, say, conventional ionization detectors;
the former are sensitive over the whole volume, implying that
surface impurities may play an important role. In principle,
bolometric detectors like those used in MiDBD, CUORICINO and CUORE
experiments are expected to have radioactive backgrounds larger
than those of conventional germanium ionization detectors, also
because the more complex technology of cryodetectors had not yet
been fully optimized from the point of view of the radiopurity of
the components near the detector. On the other hand, the
production of radiopure absorbers for thermal detectors is less
mastered than the production of radiopure Ge crystal. However,
after a considerable R\&D effort a significant improvement in the
radiopurity of bolometers has been accomplished during recent
years, and now these devices have achieved very competitive
backgrounds
\cite{Abusaidi:2000wg,Benoit:2001zu,cresst,Alessandrello2,MiDBDfinale}.

The MiDBD background \cite{Giulani} at low energy ($\sim$2
c/keV/kg/d at threshold -10 keV- and at 3500 m.w.e.) is similar to
the measured event rate anticoincident with the veto in the CDMS
experiment \cite{Abusaidi:2000wg} ($\sim$ 2 c/keV/kg/d at 10 keV
and at 20 m.w.e.) and in the EDELWEISS experiment
\cite{Benoit:2001zu}($\sim$ 1.8 c/keV/kg/d at 30 keV and 4000
m.w.e.), without a veto, (in both cases, obviously, prior to
charge-heat discrimination) and is roughly equal to that of DAMA
\cite{Bernabei} and ANAIS \cite{anais} ($\sim$ 1.5 c/keV/kg/d, at
2 keV and 3500 m.w.e. and at 4 keV and 2450 m.w.e. respectively)
but still one order of magnitude worse than that of IGEX
\cite{IGEXDM} ($\sim$ 0.2 and 0.05 c/keV/kg/d at 4 keV and 10 keV
respectively, and at 2450 m.w.e.). In the double beta decay region
($\sim$ 2.5 MeV) the MiDBD background values ($\sim$ 0.3 and 0.6
c/keV/kg/y in the old and new set-up respectively) are competitive
but still higher than those of the IGEX and Heidelberg-Moscow
experiments \cite{IGEX2beta,HM2beta} ($\sim$ 0.05 c/keV/kg/y at 2
MeV) (which uses Pulse Shape Discrimination). The challenge of
CUORE is to significantly reduce the MiDBD background values (both
in the low and high energy regions) by, say, two orders of
magnitude without using background discrimination mechanisms like
the simultaneous measurement of charge (or light) and heat. This
reduction will be accomplished by working on both the selection of
low contamination materials and on the improvement of the detector
geometry and the shielding efficiency. As one of the main
ingredients in the prospective physics potential of the CUORICINO
and CUORE experiments is the level of background achievable, an
evaluation of the expected background must be done, including the
contribution of all the possible background sources. A preliminary
estimation is contained in  Ref.~\cite{NIMpaper}, a more complete
estimate is underway. In the following we will give a briefing of
that reference and will add further considerations.

There are several background sources to be considered: the
environmental backgrounds of the underground site (neutrons from
the rocks, environmental $\gamma$  flux...), the intrinsic
radioactivities of the detectors components and shielding (bulk
and surface), the cosmic muon induced backgrounds (neutrons, muon
direct interactions \dots), as well as the possible cosmogenic
induced activities produced when the detector and components were
outside the underground laboratory. Even the small leakage of two
neutrino double beta decay counts emitted by the absorber's nuclei
in the relevant region of analysis might be a potential source of
background. The depth at which the experiment will be performed,
plus the addition of an efficient cosmic veto and a suitable
passive shielding (lead, polyethylene, \dots) will effectively
reduce the external background. So we will refer mainly to the
intrinsic background, although we will make some remarks
concerning the other sources. On the other hand, the experience
gained in the MiDBD experiment, where the measured background is
used as a test-bench for checking the MC estimates, has been
essential in predicting the expectations for CUORICINO and CUORE.
Using the MiDBD results and the CUORICINO tests a Monte Carlo
study of the expected intrinsic background has been carried out
for the CUORICINO and CUORE set-ups. Radioactive impurities in the
bulk of the tellurite crystals, as well as in the dilution
refrigerator and surrounding shielding were assumed. Background
suppression due to the anticoincidence between crystals (which
will be significant in the CUORE set up) was calculated. The code
is based on the GEANT-4 \cite{geant} package; it models in detail
the shields, the cryostat, the detector structure and the detector
array for the MiDBD, CUORICINO and CUORE experiments. It includes
the propagation of photons, electrons, $\alpha$ particles and
heavy ions (nuclear recoils from $\alpha$ emission) as well as
neutrons and muons. For the simulated radioactive chains
($^{238}U$ and $^{232}Th$) or radioactive  isotopes ($^{40}K$,
$^{60}Co$) alpha, beta and $\gamma$ /X  ray emmissions are
considered according to their branching ratios. The  time
structure of the decay chains is taken into  account and the
transport of nuclear recoils from alpha emissions is included.

\subsection{MiDBD results}

The 20 crystal MiDBD array was operated in two different
configurations \cite{MiDBDfinale}, the difference between the two
was that in the latter  configuration the Roman lead shield
surrounding the detectors was improved by increasing its
thickness, and that the crystals as well as their copper mounting
structures underwent a new surface treatment. The crystals were
re-polished with superpure powders while the copper was cleaned
with a chemical process. The result was an improvement in the
background measured in the second configuration (MiDBD-II) with
respect to that measured in the previous one (MiDBD-I) either in
the gamma region (below 3 MeV) and in the $\alpha$ region (between
3 and 10 MeV). The $\alpha$ peaks identified in MiDBD-I as due to
a surface U and Th contamination of the  crystals were reduced by
a factor 2. The reduction of the $\alpha$ contaminations, together
with the reduction of the $^{208}$Tl line at 2615 keV is probably
responsible of  the improvement obtained in the 2448-2556 keV
(0$\nu$2$\beta$ decay) region where the  counting rate changed
from b=0.59$\pm$0.06 c/keV/kg/y in MiDBD-I to b=0.33$\pm$0.11
c/keV/kg/y in MiDBD-II \cite{MiDBDfinale}. During MiDBD-II run, a
10 cm thick borated polyethylene shield was mounted outside the
external lead shield but no improvement in the counting rate was
obtained proving that the backgroud  was still dominated by other
radioactive sources. With the above background value the lower
limit of the neutrinoless half-life of $^{130}Te$ obtained is
$T_{1/2}^{0\nu}\geq 2.1\times10^{23}$ y, or equivalently, an upper
bound for the Majorana neutrino mass $\langle m_{\nu} \rangle
\lesssim$ of 0.9 - 5.2~eV (depending on the nuclear matrix element
calculation), which is the second-best published result
\cite{MiDBDfinale}. As far as the low energy region is concerned,
the background of MiDBD stands around 2.3 c/keV/kg/day between 10
and 50 keV and 0.5 c/keV/kg/day between 50 and 80 keV, region
where the dark matter signal is expected \cite{Giulani}. With
these background values, the exclusion plot of WIMPs interacting
coherently with Te and O is depicted as the (dashed) contour of
figure \ref{cuoricino_exclusion}.

To understand what are the main sources responsible of the MiDBD
background  the MC code developed to study CUORE background   was
used to reproduce the results obtained in the two configurations
in  which the MiDBD 20 crystal array was operated \cite{Capelli}.
The radioactive impurities used as inputs were all the
contaminants identified in the MiDBD background spectrum, through
an analysis of the $\alpha$ and $\gamma$  lines: $^{238}$U chain
in secular equilibrium, $^{238}$U chain broken at the  level of
$^{226}$Ra,  $^{214}$Bi or $^{210}$Pb, $^{232}$Th chain in secular
equilibrium, $^{40}$K and $^{60}$Co.  Either bulk contaminations
(with a uniform and isotropic distribution) and surface
contaminations (with an exponential density profile and various
depths) were considered. The details of this analysis are reported
in Ref.~\cite{Capelli}, the results obtained can be summarized as
follows: the MC simulation is able to reproduce the measured
background spectrum, allowing the identification and localization
of the most probable sources of background. It proves that a large
contribution to the counting rate in both the low energy region
and the 0$\nu$2$\beta$ region comes from  the surface
contaminations of the crystals and of the copper  structure facing
the bare crystals. In the case of the crystals, the origin of this
contamination is known: it is due to the use of contaminated
powders during the optical polishing of the crystals produced by
the Shanghai Quinhua Material Company (SQM) in China. Indeed the
re-treatment of their surfaces by means of low contamination
powders reduced the contamination to one half. For this reason it
was agreed that the CUORICINO crystals would undergo only a minor
surface treatment in China (the one indispensable to check crystal
quality) and were subsequently polished  in Italy using low
contamination powders.  On the contrary, the origin of the copper
surface contamination is still uncertain: it could either have
been produced during the machining of the metal or during the
surface acid treatment. As the CUORICINO copper structure was
produced with the same technique of the MiDBD-II a similar
contamination may have been produced. For CUORE a dedicated R$\&$D
is foreseen  to reduce or cancel this kind of contamination.
Finally, other important contributions come from bulk
contamination of the structures outside the Roman lead shield that
surrounds the detector and from outside the cryostat. Evaluating
each impurity for each simulated element, the maximum contribution
to the measured background by bulk contamination of TeO$_2$,
copper and Roman lead were obtained and have been used in CUORE
simulation.

\subsection{CUORICINO and CUORE instrinsic background}

CUORICINO will be operated in the same set-up as that of the MiDBD
experiment: the same dilution refrigerator, the same cryostat, the
same external lead and borated-polyethylene shield, at the same
underground location (a description of the set-up can be found in
Ref.~\cite{MiDBDfinale}). The crystals will be different but the
measured upper limits of their U and Th contamination are again of
the order of 10$^{-13}$~g/g . Therefore bulk contribution to
CUORICINO background should be not larger than that  measured in
MiDBD-II. On the other hand, for surface contamination some
improvement with respect to MiDBD-II result is foreseen thanks to
the different surface treatment of the crystals and to a reduced
exposure of detectors to dust and Rn. A reasonable guess for
CUORICINO is therefore a background of $\sim$ 0.1 c/keV/kg/y in
the $0\nu2\beta$ region, slightly better than the one measured in
MiDBD-II; while in the dark matter region a background of the
order of 1 - 0.1 c/keV/kg/d is anticipated.

On the contrary no extrapolation of the background achievable in
CUORE is possible  from the MiDBD data, nor from the CUORICINO
results.  This is true for several reasons: the different
structure of the detector (in the case of CUORE  an efficient
reduction of background events will be provided by operating the
detectors  of the array in anticoincidence, without loosing
efficiency in detecting WIMPs or 0$\nu$2$\beta$ decay events), the
different cryostat and shielding systems, and finally because a
much more severe selection of the constructing materials and a
much better control on  their mechanical working and chemical
cleaning will be applied.  Also predictions based on MC
simulations are not straightforward since they depend critically
on the assumed inputs: contamination levels, detector design, etc.
Here we present a very conservative evaluation of the background
reachable with CUORE, mainly based on the state of the art of
detector design and on the knowledge of radioactive contamination.
The CUORE construction will require about five years and in the
meantime  an $R\& D$ dedicated to background reduction will be
realised. This approach for intrinsic  background  evaluation
could be rather pessimistic but it is the only one that  presently
guarantees a reliable but possibly conservative prediction.

In CUORE (a description of the set-up can be found in
Ref.~\cite{NIMpaper}) the main contribution to background due to
bulk contaminations comes from the heavy  structures near the
detectors (the copper mounting structure of the array,  the
Roman-lead box and the two lead disks on the top of the array) and
from the detectors  themselves (the $TeO_2$ crystals). In the
simulation, the $^{232}Th$, $^{238}U$, $^{40}K$, $^{210}Pb$, $^{134}Cs$, $^{137}Cs$ and $^{207}Bi$
impurities of $TeO_2$, copper and lead as well as the
$^{60}Co$ cosmogenic contamination of copper are conservatively
assumed equal to the  90\% C.L. upper limits obtained for the
contamination levels of those materials measured in  MiDBD
experiment or in low HP$Ge$ spectrometry \cite{Capelli,NIMpaper}.
The $^{60}Co$ cosmogenic contamination of $TeO_2$ is on the other
hand evaluated according to the time required to produce $TeO_2$
powder from the ore, grow the crystals and store them underground
\cite{COSMO}.  With these contaminations, assuming a threshold of
10 keV and using the results obtained after the  reduction by
anti-coincidence between detectors, the background due to bulk
contaminations is $\sim 4\times10^{-3}$~counts/(keV~kg~y) in the
$0\nu 2\beta$ decay region and $\sim
3\times10^{-2}$~counts/(keV~kg~d) in the low energy region
(10-50~keV). On the other hand, surface contaminations contribute
to background only when they are localized on the  crystals or on
the copper mounting structure directely facing them. In the
simulation, U and Th  surface contaminations are assumed to be:
for crystals $\sim$~100 times lower and for copper $\sim$~50 times
lower than the corresponding contamination in MiDBD experiment.
The MiDBD crystals were contaminated during the polishing
procedure due to the use of highly  contaminated powders.
Polishing powders with a radioactive  content 1000 times lower are
however commercially available and have already been used  for
CUORICINO, an improvement of the surface contamination of a factor
100 is therefore fully justified. A similar situation holds for
copper. In the MiDBD experiment the copper surfaces were treated
with an etching procedure studied to reduce impurities on surfaces
before the  sputtering process.  This procedure resulted in an
improvement of the surface quality of copper, however it is not
optimized from the point of view of background. The use -for the
surface treatment- of low contamination liquids in a low
background environment, and special care in the mechanical working
and handling will allow an improvement between one and two orders
of magnitude in the surface contamination of copper. With such a
contamination the surface contribution is estimated to be
2.8$\times 10^{-3}$~counts/(keV~kg~y) in the 0$\nu$2$\beta$ region
and $1.2\times 10^{-3}$~counts/(keV~kg~d) in the dark matter
region.

Finally, the unavoidable background produced by the $2\nu2\beta$
decay region is lower than  10$^{-4}$~counts/(keV~kg~y) in the
0$\nu$2$\beta$ region and $\sim 10^{-4}$~counts/(keV~kg~d) in the
dark matter region  (assuming 3.8$\times10^{20}$ y for the
$^{130}$Te $2\beta2\nu$ half-life \cite{MiDBDfinale}).

\subsection{Cosmogenic activation}

Concerning the cosmogenic activation produced by cosmic rays when
the crystals are above ground (during fabrication and shipping of
the crystals from the factory to the underground laboratory) we
used a code based on computed cross sections to estimate the
amount and type of radionuclides produced by cosmic rays on
TeO$_2$ \cite{COSMO}.  The comparison between our prediction and
the MiDBD-I and II run data is discussed in ref.~\cite{Capelli}.
The radionuclei produced by the activation of tellurium are mostly
tellurium isotopes (A=121,123,125,127) as well as $^{124}$Sb,
$^{125}$Sb,  $^{60}$Co and tritium, these last three being of more
concern because of their long half-life ($^{125}$Sb:~beta decay of
2.7 years, end-point energy of 767 keV, $^{60}$Co:~beta decay of
5.27 years end-point energy 2823 keV and $^{3}$H:~beta decay of
12.3 years, end-point energy of 18 keV). In MiDBD-I no clear
evidence of cosmogenic  activation was present (the crystals were
stored underground for several months), while in MiDBD-II the
$\gamma$  lines of the tellurium isotopes were clearly seen, with
an intensity that is in good agreement with our prediction (based
on the period they were reexposed because of the repolishing
process).

The history of the CUORICINO crystals is rather similar to those
of MiDBD, therefore their cosmogenic activation is already
included in the extrapolation of CUORICINO background from MiDBD
data. In the case of CUORE the control on crystal production will
be more severe. The SQM that produces the crystals, oxidizing the
tellurium ore and growing a 760 g crystal in about two months,
will require two years to grow the 1000 CUORE detectors. Once
grown, the crystals will be shipped to Italy and stored
underground, therefore their  total exposition to cosmic rays will
be limited to about 4 months.  The total induced activities
remaining after 2 years underground have been estimated
\cite{COSMO} and the consequent contribution to the detector
counting rate was deduced by a MC simulation. The radionuclei that
contribute to background through their $\beta^-$ decay are:
\begin{itemize}
\item{in the 0$\nu$2$\beta$ region, the long lived $^{60}$Co isotope with an activity of   $\sim$0.2~$\mu$Bq/kg, while  a minor contribution is due to the isotopes $^{110m}$Ag and $^{124}$Sb whose activity is 4 times lower and fast decreasing with time;}
\item{in the dark matter region, the long lived nuclei of $^{3}$H and $^{125}$Sb with an activity of  $\sim$7~$\mu$Bq/kg for the former and of $\sim$15~$\mu$Bq/kg for the latter.}
\end{itemize}
The influence of $^{60}$Co in the CUORE background was already
considereded in the evaluation of the contributions due to bulk
contaminations of the crystals, while the contribution of $^{3}$H
and $^{125}$Sb to the dark matter region is completely negligible
($\sim 10^{-3}$~counts/(keV~kg~d)) if compared to the intrinsic
background.

\subsection{Underground neutron, $\mu$ and $\gamma$ interactions}

As noted before, neither contributions from underground cosmic
muons nor neutrons have been taken into account in detail in the
estimation of the background. However, the following simplified
arguments will serve to have an approximate idea of their
contribution.  The depth of the LNGS (3500 m.w.e) reduces the muon
flux to $\sim$2$\times 10^{-8}$~cm$^{-2}$s$^{-1}$, but a further
effective reduction could be obtained with the use of an efficient
(99.9\%) active veto to detect muons traversing the detectors and
to tag possible events associated with them. Consequently, the
muon-induced events contributing to the background are expected to
be only a small component of it. On the other hand, the shielding
will substantially reduce the event rate due to particles external
to the detector from various sources (neutrons and photons), from
radioactivity in the environment (natural decay chains U/Th,
$^{210}$Pb, $^{40}$K, \dots), as well as muon-induced in the
surroundings or in the shielding itself. The passive shielding
either for MiDBD, for CUORICINO and for CUORE consists of a
neutron screen (blocks of -borated- polyethylene  of 10 cm
thickness) to moderate and attenuate neutrons and  a lead shell to
attenuate the incoming external photons; the innermost part (10 cm
thick) is made of special lead with a small content ($<$16 Bq/kg)
of $^{210}$Pb (half-life 22 years)  while the outer part (10 cm
thick) is made of modern lead  ($^{210}$Pb $<$150Bq/kg) both with
a low level of Th and U.

Neutrons may constitute a worrisome background in dark matter
experiments because for appropriate neutron energies (few MeV)
they can produce nuclear recoils in the detector target nuclei
which would mimic WIMP interactions. Simple kinematics tells that
in the case of tellurium, neutrons of 1 to 5 MeV could elastically
scatter off tellurium nuclei producing recoils of energies up to
31 to 154 keV. In general, one considers neutrons of two origins:
from radioactivity in the surroundings or muon-induced. Depending
on the overburden of the underground site (i.e., depending on the
muon flux), muon-induced neutrons are produced, at lesser or
greater rate, both inside and outside the shielding. They are
moderated according to their energies by the polyethylene and lead
shield (when produced outside) or tagged by the muon veto
coincidence (when produced within the passive shielding). A
certain fraction of the neutrons of a few MeVs produced outside
can pass the veto reaching the detector and producing "dangerous"
nuclear recoils and $\gamma$ background. However, a significant
fraction of the associated events can be vetoed because of their
interaction with the veto and also because of the hadronic showers
initiated by the muons (see Ref. \cite{Abusaidi:2000wg}).

In the case of external neutrons (from the rocks, from fission
processes or from (n,$\alpha$) reactions, as well as neutrons
originated by muons in the walls of the underground site), the
environmental neutron flux has been measured in LNGS. The
resulting flux is of $\sim1\times 10^{-6}$ cm$^{-2}$s$^{-1}$ for
the thermal component, $\sim2\times 10^{-6}$ cm$^{-2}$s$^{-1}$ for
the epithermal and $\sim2\times 10^{-7}$ cm$^{-2}$s$^{-1}$ for
energies over 2.5 MeV \cite{Belli}. They are fairly well moderated
by the polyethylene and eventually absorbed or captured.  We have
carried out a Monte Carlo simulation of the propagation of
neutrons through the 10 cm thick borated polyethylene shield of
CUORICINO and CUORE. The result is that the neutron  induced event
rate on the entire energy range (from threhsold to 10 MeV) is much
lower  than the contribution due the bulk contamination of
crystals.  Other neutrons, produced by muon interactions inside
the shielding materials, are very scarce and tagged as events
coincident with the muon veto. As it is well-known, muon-induced
neutrons are originated in a variety of processes. The reduction
of muon flux in underground sites results in a substantial
depletion of the associated neutrons and below $\sim$100 m.w.e.
the dominant sources of neutrons are nuclear fission processes and
(n,$\alpha$) reactions in rocks and other environmental material
with sizeable content of U/Th. The energy spectrum of muon-induced
neutrons is approximated by an inverse energy power law
($E^{-0.88}$ for 1-50 keV and $E^{-1}$ above 50 keV) but other
neutron energy spectra have been proposed.
The neutron yield per muon can be approximately evaluated for an organic compound through the simple expression $N_{n}=4.14\times E_{\mu}^{0.74} \times 10^{-6}$ neutrons/(g~cm$^{-2}$) per muon \cite{Wang} which fits the value of 1.5$\times 10^{-4}$ neutrons/(g~cm$^{-2}$) of the muon-induced neutron flux measured by the LVD experiment at Gran Sasso \cite{Aglietta}.

For the LNGS muon flux (2.5$\times 10^{-8}$ $\mu$/(cm$^{2}\,$s)),
muons would produce in the CUORE shielding of polyethylene (10 cm)
and lead (20 cm)) about $\sim$0.04 neutrons/(m$^2$ day) in the
polyethylene shield and $\sim$25 neutrons/(m$^2$ day) in the lead
shell.

So, independently of the mechanism set up to reject or tag the events associated to neutrons, their rather small number is expected to play a secondary role in the total background compared with other, intrinsic, sources of background.

A preliminary evaluation of the influence of the environmental
$\gamma$ background in Gran Sasso  \cite{Arpesella} resulted in a
negligible contribution for the $0\nu2\beta$ region and a
contribution similar to that of bulk contaminations for the dark
matter region.

A more complete and detailed study of the external, non intrinsic background rates for CUORE is underway.

\subsection{Expected performances for CUORICINO and CUORE}

Summarizing, for CUORICINO a background similar to the one
measured in MiDBD-II is the most conservative prediction either in
the dark matter region or in the $0\nu2\beta$ region. For CUORE on
the other hand, the previous sections have shed light on how the
background contamination dominating the two regions of interest is
mostly the intrinsic background due to bulk and surface
contaminations of the contructing materials. With the presently
achieved quality of low  contamination materials and considering
the worst possible condition for bulk contaminations (i.e. all the
contamination equal to the present 90\%/,C.L. measured upper
limits) we have proved that CUORE background will be $\sim
0.007$~counts/(keV~kg~y) at the $0\nu2\beta$ transition energy and
$\sim 0.05$~counts/(keV~kg~d) near threshold.  Even at these
conservative values CUORE is, as we will prove in the following, a
powerful instrument to search for double beta decay and dark
matter. However, the goal of the CUORE technical $R\,\& D$ of the
next few years will be the reduction of background to the level of
0.001~counts/(keV~kg~y) at the $0\nu2\beta$ transition energy and
to 0.01~counts/(keV~kg~d) at threshold.  The potential of the
experiment in this background configuration will also be
discussed.

Regarding the expected threshold and resolution, in the CUORICINO
test experiment \cite{Pirro:2000fi} energy thresholds of $\sim 5$
keV have been obtained and a resolution of 1 keV at the 46 keV
line of $^{210}$Pb was achieved in some of the detectors. Details
can be found in Ref. \cite{Giulani}. We will assume a conservative
value of 10 keV for the energy threshold of both CUORICINO and
CUORE, and energy resolutions of 1 keV at threshold. As far as the
energy resolutions obtained in the double beta decay region,
values of 3 keV at 2615 keV were achieved in some crystals but
they are worse (by a factor two) in others. An energy resolution
of 5 keV will be assumed both for CUORE and for CUORICINO. Taking
into account these expectations, we discuss in the following the
prospects of CUORICINO and CUORE for double beta decay searches
(section 3), for WIMP detection (section 4) and for solar axion
exploration (section 5).

\section{Double beta decay} \label{dbd}

One of the main scientific objectives of the CUORE detector is the
search for the neutrinoless double-beta decay of the $^{130}$Te
isotope contained in the (natural) TeO$_2$ crystals.

\subsection{Theoretical motivation}

The importance of nuclear double-beta decay as an invaluable tool
to explore particle physics beyond the Standard Model has been
repeatedly emphasized and widely reported \cite{doblebeta}. In the
Standard Model of Particle Physics neutrinos are strictly
massless, although there is no theoretical reason for such a
prejudice. On the experimental side, moreover, there exist strong
evidences from atmospheric neutrino data (from SuperKamiokande
\cite{fukuda}), from experiments with solar neutrinos (from
Homestake, Gran Sasso and Kamioka \cite{hampel}) and from reactor
experiments (from KamLAND \cite{kamland}) which suggest that
neutrinos have indeed masses and oscillate among the three
species. The results of the solar $\nu$ experiment from SNO with
both CC (Charged Current) and NC (Neutral Current) interactions
\cite{Ahmad:2001an}, also combined with SuperK have definitively
provided a strong evidence that neutrinos change of flavour and,
consequently, the existence of non-zero mass neutrinos. The recent
results of KamLAND \cite{kamland} exclude all oscillation
solutions but the Large Mixing Angle solution to the solar
neutrino problem using reactor antineutrino sources. However,
neutrino oscillation experiments provide the squared mass
difference between the neutrino species but not their absolute
value and scale. The neutrinoless double beta decay would help to
solve this question and to disentangle the hierarchy scheme of the
neutrino flavours. Most of the models (see \cite{pascoli})
indicate that the Majorana neutrino mass parameter could be around
(or slightly below) $\langle m_{\nu}\rangle \sim 0.05$ eV, value
within reach of the future double beta decay experiments, like
CUORE. Majorana massive neutrinos are common predictions in most
theoretical models, and the value of a few tens of
millielectronVolts predicted for its effective mass, if reached
experimentally -as expected- will test its Majorana nature.
Double-beta decay experiments with even better sensitivities (of
the order of a few meV) will be essential to fix the absolute
neutrino mass scale and possibly to provide information on CP
violation \cite{pascoli,feruglio}. On the other hand, galaxy
formation requires a small amount of hot non-baryonic dark matter
likely in form of neutrinos to match properly the observed
spectral power at all scales of the universe. The question of the
neutrino mass is one of the main issues in Particle Physics.

In the Standard Model, neutrinos and antineutrinos are assumed to
be different particles, but no experimental proof has been
provided so far. Nuclear double-beta decay addresses both
questions: whether the neutrinos are self-conjugated and whether
they have Majorana mass. In fact, the lepton number violating
neutrinoless double beta decay $(A,Z)\rightarrow(A,Z+2)+2e^-$
($2\beta 0\nu$) is the most direct way to determine if neutrinos
are Majorana particles. Another form of neutrinoless decay,
$(A,Z)\rightarrow(A,Z+2)+2e^- + \chi$ may reveal also the
existence of the Majoron ($\chi$), the Goldstone boson emerging
from the spontaneous symmetry breaking of the $B-L$ symmetry which
is of most relevance in the generation of Majorana neutrino mass
and has far-reaching implications in Astrophysics and Cosmology.
Moreover, the observation of a $2\beta 0\nu$ decay would imply a
lower bound for the neutrino mass, i. e. at least one neutrino
eigenstate has a non-zero mass.

It is quite straightforward to demonstrate how the determination of the effective Majorana mass of the
electron neutrino constrains the lightest neutrino mass eigenvalue. This was done recently by Barger,
Glashow, Marfatia and Whistnant \cite{Barger}. However it is very convenient to use the approximation that
$sin\,\theta_{2} \simeq 0 \equiv s_{2}$ and $cos\,\theta_2 \simeq 1$. With the further assumption
$\delta m_{\odot}^{2} \ll \delta m_{AT}^{2}$ in the normal hierarchy case we obtain:
\begin{equation}
|< m_\nu >| \,\le\, m_1\, \le\, \frac{|< m_\nu >|}{cos(2\theta_3)}
\end{equation}
where $cos(2\theta_3)\simeq 0.5$.
Similarly in the case of inverted hierarchy:
\begin{equation}
\sqrt{|< m_\nu >|^2-\delta m_{AT}^{2}}\, \leq\, m_1\,\leq\,
\frac{\sqrt {|< m_\nu >|^2-\delta m_{AT}^{2}cos(2\theta_3)}}{cos(2\theta_3)}
\end{equation}
A related quantity, $\Sigma \equiv m_1+m_2+m_3 $ has great importance in cosmology. It is related to the
quantity $\Omega_{\nu}$, the fraction of the critical density, $\rho_0$ (that would close the universe) that
is in the form of neutrinos, where:
\begin{equation}
\Sigma = (93.8\,eV)\; \Omega_{\nu}\;h^2
\end{equation}
and $h$ is the dimensionless Hubble constant. It was shown in reference \cite{Barger} that $\Sigma$
obeys the
following inequality with respect to $|< m_\nu >|$:
\begin{eqnarray}
\lefteqn {2\,|< m_\nu >|+\sqrt{|< m_\nu >|^2 \pm \delta m_{AT}^{2}} \leq \Sigma} \nonumber\\
& & \leq \frac{2\,|< m_\nu >|+\sqrt {|< m_\nu >|^2 \pm \delta m_{AT}^{2}\,cos(2\theta_3)}}{cos(2\theta_3)}
\end{eqnarray}
where the $+$ and $-$ signes refer to normal and inverted
hierarchies respectively, and $cos(2\theta_3) \equiv 0.5$. If this
relation is written as two equalities, and each side is solved for
$|< m_\nu
>|$ in terms of $\Sigma$, quadratic equations result with the
constant terms $(\Sigma^2 \mp \delta m_{AT}^{2})$. Values of
$\Sigma^2$ of cosmological interest are much larger than $\delta
m_{AT}^{2} \leq 0.005$ eV (99.73\% C.L.). Accordingly the above
inequality reduces to the trivial form:
\begin{equation}
|< m_\nu >|\, \leq \,\Sigma / 3\, \leq \, 2|< m_\nu >|
\end{equation}
This clearly demonstrates the direct connection between neutrinoless double beta decay and neutrino dark
matter.
Cosmic microwave background and galaxy cluster surveys have set the bound $\Sigma \leq 1.8\,eV$ \cite{ergaroy}.
It is expected that the MAP satellite will produce data that will result in a sensitivity $\Sigma \simeq
0.5\,eV$.

One possible interesting scenario results in the case the MAP
satellite, if for example it observes a clear singal well above
their limiting sensitivity, and CUORE (as well as other next
generation experiments) have a negative results at sensitivities
of $\Sigma$ far below that of the satellite data. This would be a
clear indication that the phenomenon causing the density
fluctuations implied by the CMB data was not caused by Majorana
neutrino dark matter. In addition a positive measurement of the
mass $m_{\nu_e}$ by KATRIN $^3 He$ beta end-point spectrum
measurement would have to yield $m_{\nu_e} \geq 0.35\,eV$. The
CUORE experiment will have a far greater sensitivity, and if it
found a negative result, the mistery of Dirac or Majorana
character of neutrinos will be solved.

These and other issues, make the search for the neutrinoless
double beta decay an invaluable tool for the exploration of
non-standard model physics, probing energy scales well above those
reachable with accelerators. That is the motivation of why there
are dozens of experiments underway looking for the double beta
decay of various nuclei \cite{doblebeta} like $^{76}Ge$ (IGEX
\cite{IGEX2beta}, Heidelberg-Moscow \cite{HM2beta}), $^{100}Mo$
and others (NEMO \cite{sarazin}, ELEGANTS \cite{elegants}) and
$^{130}Te$ (MiDBD, CUORICINO) and a few big experimental projects,
like CUORE, Majorana \cite{majorana} ($^{76}Ge$), MOON
\cite{ejiri} ($^{100}Mo$) and EXO \cite{danilov} ($136Xe$).

\subsection{Experimental prospects}

The cryogenic thermal detectors provide new double-beta emitter
nuclei to be explored in "active" source/detector calorimeters.
Some of them have been tested and others are already in running
detectors, like $^{48}$Ca in CaF$_2$, $^{130}$Te in TeO$_2$, and
$^{116}$Cd in CdWO$_4$. As far as the Tellurium Oxide is
concerned, the $^{130}$Te isotope is a good candidate for double
beta decay searches: its isotopic content in natural Tellurium is
33.87\%, and its 2$\beta$ Q-value ($Q_{2\beta}=2528\pm1.3$ keV) is
reasonably high to escape from the main radioimpurity lines when
looking for a neutrinoless signal. Moreover, this Q-value happens
to be between the peak and the Compton edge of the 2615 keV line
of $^{208}$Tl, which leaves a clean window to look for the signal.
Finally, it has a fairly good neutrinoless nuclear factor-of-merit
$F_N^{0\nu}=G_{0\nu}\left|M^{0\nu}\right|^2$, where $G_{0\nu}$ is
an integrated kinematical factor qualifying the goodness of the
$Q_{2\beta}$ value (large phase space) and $M^{0\nu}$ is the
neutrinoless double-beta decay nuclear matrix elements
characterizing the likeliness of the transition.

In table \ref{tab_doblebeta} we quote the values of $F_N^{0\nu}$
in the case of $^{130}$Te calculated in various nuclear models
\cite{doblebeta}, together with those of other emitters used in
source/detector calorimeters. It can be seen that no matter the
nuclear model is used to compute the neutrinoless double-beta
decay matrix elements, the merits of $^{130}$Te are a factor 5--10
more favorable than those of $^{76}$Ge (the emitter where the best
neutrinoless double beta decay half-life limits have been achieved
so far), which translates into a factor 2 to 3 better as far as
the $\langle m_{\nu} \rangle$ (Majorana neutrino mass parameter)
bounds are concerned.

The detector factor-of-merit $F_D^{0\nu}$, or detection
sensitivity, introduced earlier by Fiorini, provides an
approximate estimate of the neutrinoless half-life limit
achievable with a given detector. For source/detector devices, it
reads:

\begin{equation}\label{merit}
F_D^{0\nu} = 4.17\times 10^{26} \times \frac{a}{A}
\sqrt{\frac{Mt}{b\Gamma}}\times \epsilon {\rm \ \ \ years}
\end{equation}

\noindent where $A$ is the atomic mass, $a$ is the isotopic
abundance, $M$ the detector mass in kg, $b$ the background in
c/keV/kg/y in the 2$\beta$ neutrinoless decay region, $t$ the
running time in years, $\Gamma$ the FWHM energy resolution in keV
and $\epsilon$ the detector efficiency (which is 0.86 in the case of the cubic 5x5x5 cm$^3$ crystals). In the case of a
TeO$_2$ crystal detector, $F_D^{0\nu} \sim 7.59 \times 10^{23}
\sqrt{\frac{Mt}{b\Gamma}}$, with $M$ the crystal mass in kg and
$b$ the background in counts per keV and year per kg of detector
mass.

A simple projection for CUORICINO sensitivity can be obtained
assuming a background of b=0.1 c/keV/kg/y (as discussed in Section
2) and a resolution of $\Gamma=$ 5 keV in the 2.5 MeV region. In
that case, one would have $F_D\sim 1.07\times 10^{24} \sqrt{Mt}$
years, with $M$ (kg) the mass of the TeO$_{2}$ crystal array. For
the mass of CUORICINO (M=40.7 kg), one will have a sensitivity of
$F_D\sim 6.86\times10^{24} \sqrt{t}$ years. Using a typical
average value of $F_{N}=4\times 10^{-13}$ y$^{-1}$, as obtained in
QRPA \cite{QRPA1,QRPA2,QRPA3,QRPA4,QRPA5}, CUORICINO will have a
mass sensitivity of $\langle m_{\nu} \rangle <0.3$ eV in one year,
in the least favourable case. Notice that the bound on the
effective mass of the electron neutrino, $\langle m_{\nu}
\rangle$, is given by $\langle m_\nu \rangle \leq m_e /
\sqrt{F_D^{0\nu}F_N^{0\nu}}$. Using for comparison the value of
$F_{N}=5.33\times 10^{-13}$ y$^{-1}$ of Ref. \cite{QRPA1} which is
usually employed in the $\langle m_{\nu} \rangle$ bound
($\sim$0.33 eV) derived from the Ge experiments, with these
assumptions CUORICINO would provide $\langle m_{\nu} \rangle
<$0.24 eV.

To go further, one needs to increase the mass of TeO$_{2}$ and to
reach even a lower background. In our very conservative projection
for the CUORE array (760 kg of TeO$_{2}$) a background of b=0.01
c/keV/kg/y can be easly achieved, assuming as in the case of
CUORICINO a resolution $\Gamma$(2.5 MeV)=5 keV, one would get
$T_{1/2}^{0\nu}\geq 9.4\times10^{25}\sqrt{t}$ years, which in one
year of statistics would provide $\langle m_{\nu} \rangle$ bounds
ranging from 0.07 eV \cite{QRPA1}, 0.08 eV \cite{QRPA3}, 0.15 eV
\cite{QRPA5} or 0.04 eV \cite{WCSM} just to mention results using
a few nuclear matrix element estimates. However, the R\&D to be
carried out in CUORE, if succesful, would provide a value of
$b\sim 0.001$ c/keV/kg/y i.e., a detection sensitivity of $F_D\sim
2.96\times10^{26} \sqrt{t}$ years, or $\langle m_{\nu} \rangle$
bounds ranging from $\sim 0.05~t^{-1/4}$ eV (in
\cite{QRPA1,QRPA2,QRPA3,QRPA4,QRPA5,OEM}) to $\sim 0.03~t^{-1/4}$
eV (in \cite{WCSM,GenSen}). TeO$_{2}$ crystals made with $^130$Te
enriched material have been already tested during the MiDBD
experiment, making an enriched CUORE a feasible option. Assuming a
95\% enrichment in $^130$Te and a background level of b=0.001
c/keV/kg/y, the sensitivity becomes $F_D\sim 8.32\times10^{26}
\sqrt{t}$ years. For an exposure of 5 years, the corresponding
$\langle m_\nu \rangle$ bounds range from 9 meV to 56 meV
depending on the nuclear matrix element calculations.

\section{WIMP detection} \label{wimps}

Recent cosmological observations \cite{Bachcall:1999} provide
compelling evidence for the existence of an important component of
non-baryonic cold dark matter in the Universe. Among the
candidates to compose this matter, Weakly Interacting Massive
Particles (WIMPs) and axions are the front runners. The lightest
stable particles of supersymmetric theories, like the neutralino
\cite{Jungman:1996df}, constitute a particular class of WIMPs.

Under the hypothesis that WIMPS are the main component of the dark
matter, these particles should fill the galactic haloes and
explain the flat rotation curves which are observed in many
galaxies. The detection of such particles could be attempted both
by means of direct and indirect methods. The direct detection of
WIMPs relies on the measurement of their elastic scattering off
the target nuclei of a suitable detector\cite{Mor99}. The non
relativistic and heavy (GeV -- TeV) WIMPs could hit a detector
nucleus producing a nuclear recoil of a few keV. Because of the
small WIMP-matter interaction cross sections the rate is extremely
low. In the case of SUSY WIMPs, most of the cross section
predictions \cite{Ellis,Bottino:2001jx,Bergstrom} (derived using
MSSM as the basic frame implemented with different unification
hypothesis) encompass a range of values several orders of
magnitude (the so-called scatter plots) providing rates ranging
from 1 c/kg/day down to 10$^{-5}$ c/kg/day according to the
particular SUSY model.

It is well known that the predicted signal for the WIMP elastic
scattering has an exponentially decaying energy dependence, hardly
distinguishable from the background recorded in the detector. The
simple comparison of the theoretical WIMP spectrum with the one
experimentally obtained, provides an exclusion curve (at a given
confidence level), as dark matter component of the halo, of those
WIMPs with masses ($m$) and cross sections on nucleons ($\sigma$)
which yield spectra above the measured experimental rate. To claim
a positive identification of the WIMP, however, a distinctive
signature is needed. The only identification signals of the WIMP
explored up to now are provided by the features of the Earth's
motion with respect to the dark matter halo. In particular, the
annual modulation \cite{Drukier:1986tm} is originated by the
combination of the motion of the solar system in the galactic rest
frame and the rotation of the Earth around the Sun. Due to this
effect, the incoming WIMP's velocities in the detector rest frame
change continuously during the year, having a maximum in summer
and a minimum in winter. Therefore the total WIMP rate changes in
time with an oscillating frequency which corresponds to an annual
period and a maximum around the beginning of June.

The relative annual variation of the signal is small (a few
percent) so in order to detect it one needs large detector masses
to increase statistics and several periods of exposure to minimize
systematics. Several experiments have already searched for this
effect \cite{modza,modarg,damaxe} and since 1997 one group has
reported a positive signal \cite{Bernabei} which has been
appearing throughout four yearly periods. The present situation is
no doubt exciting: on one hand that result has triggered an
intense activity in the field; on the other, the experimental
sensitivities of various types of underground detectors are
entering the supersymmetric parameter space \cite{Bottino:2001jx}
and in particular several of them
\cite{Abusaidi:2000wg,Benoit:2001zu,IGEXDM,zeplin} are excluding,
to a larger or lesser degree, the region of mass and cross-section
where the reported WIMP is supposed to exist. New data from one of
them \cite{Benoit:2001zu} have excluded totally the DAMA region.
We will discuss in the following the capabilities of CUORICINO and
CUORE to exclude or detect WIMPs using the total time-integrated
experimental rate and comparing it with the predicted nuclear
recoil rate. To look for the annual modulation signal in CUORICINO
/ CUORE experiments, which in principle have enough mass to be
sensitive to it, one needs to know their stability performances.
The analysis of the CUORE / CUORICINO potential for annual
modulation searches will be performed -following statistical
consideration- (see Ref. \cite{Cebrian:2001qk}), with the proviso
that systematic uncertainties are under control. Data on the
stability of CUORICINO will be crucial to assess such hypotheses.

To calculate the theoretical WIMP rate, standard hypothesis and
astrophysical parameters are assumed, i.e., that the WIMPs form an
isotropic, isothermal, non-rotating halo (the isothermal sphere
model) of density $\rho=0.3 $ GeV/cm$^3$, which has a maxwellian
velocity distribution with $v_{rms}=270$ km/s (with an upper cut
corresponding to an escape velocity of 650 km/s), and a relative
Earth-halo velocity of $v_r=230$ km/s). Other, more elaborated
halo models, which have been considered recently \cite{halomodels}
would lead to different results. The same applies when other
astrophysical parameters are employed or when uncertainties in the
halo WIMPs velocity distribution are included \cite{astropar}. The
theoretical predicted rate is expressed in terms of the mass and
cross-section of the WIMP-matter interaction. The cross sections
are normalized per nucleon assuming a dominant scalar interaction,
as is expected, for instance, for one of the most popular dark
matter candidates, the neutralino:

\begin{center}
\begin{equation}\label{norm}
    \sigma_{N\chi} = \sigma_{n\chi}A^2 \frac{\mu^2_{W,N}}{\mu^2_{W,n}}
\end{equation}
\end{center}

\noindent where $A$ is the target (oxygen and tellurium) mass
number, $\mu^2_{W,N}$, is the WIMP-nucleus reduced mass, and
$\mu^2_{W,n}$ the WIMP-nucleon reduced mass. The Helm
parameterization \cite{Helm} is used for the scalar nucleon form
factor. The $(m,\sigma)$ exclusion plot is then derived by
requiring the theoretically predicted signal for each $m$ and
$\sigma$ in each energy bin to be less than or equal to the (90\%
C.L.) upper limit of the (Poisson) recorded counts. The bin width
is assumed to be equal to the detector resolution.

\begin{figure}[h]
\begin{center}
\includegraphics[width=1\textwidth]{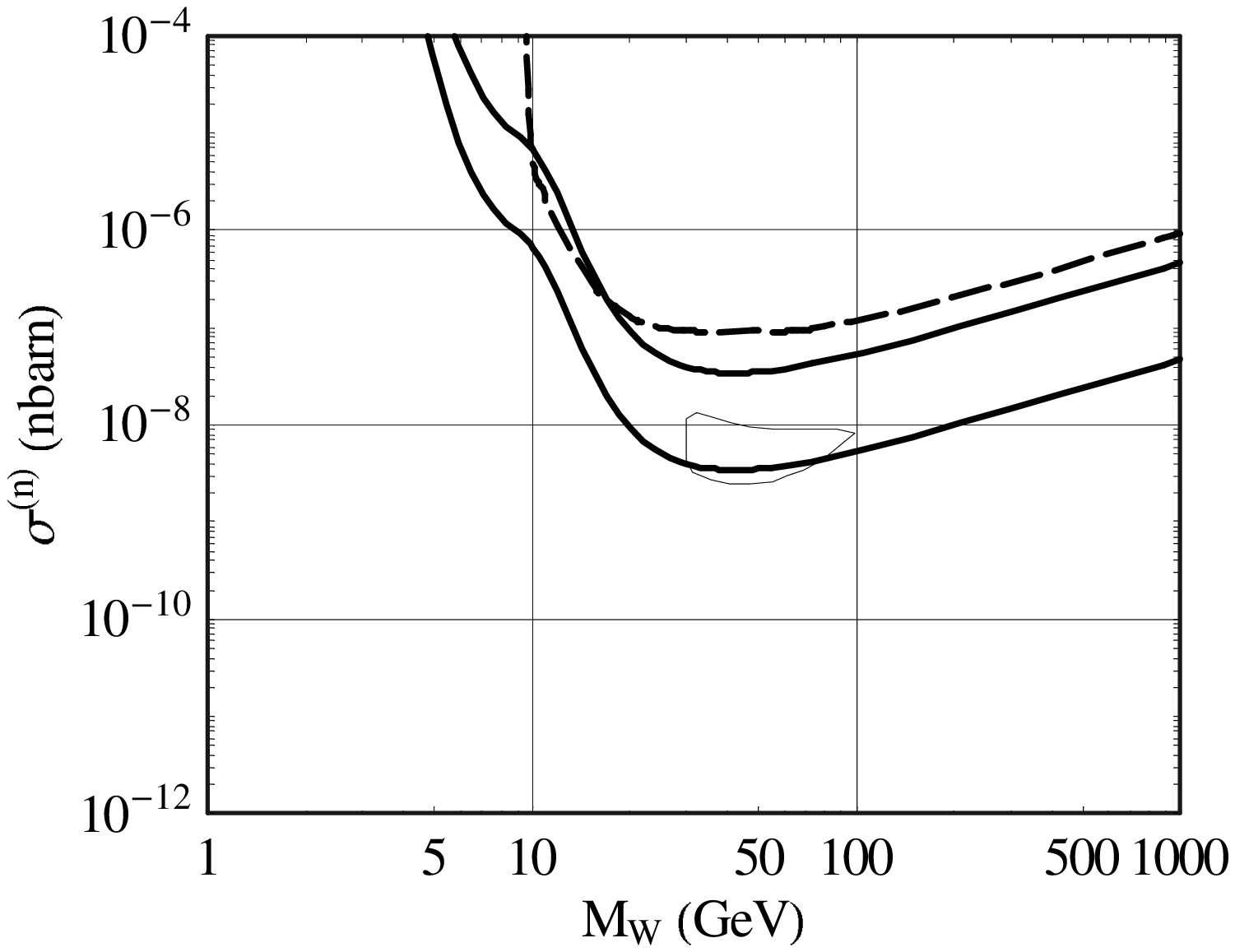}
\caption{Exclusion projected for 2 years of CUORICINO assuming a
threshold of 10 keV, a low energy resolution of 1 keV, and low
energy background levels of 1 and 0.1 c/keV/kg/day respectively.
The closed curve represents the DAMA region. The dashed line
corresponds to the current MiDBD result.}
\label{cuoricino_exclusion}
\end{center}
\end{figure}

\begin{figure}[h]
\begin{center}
\includegraphics[width=1\textwidth]{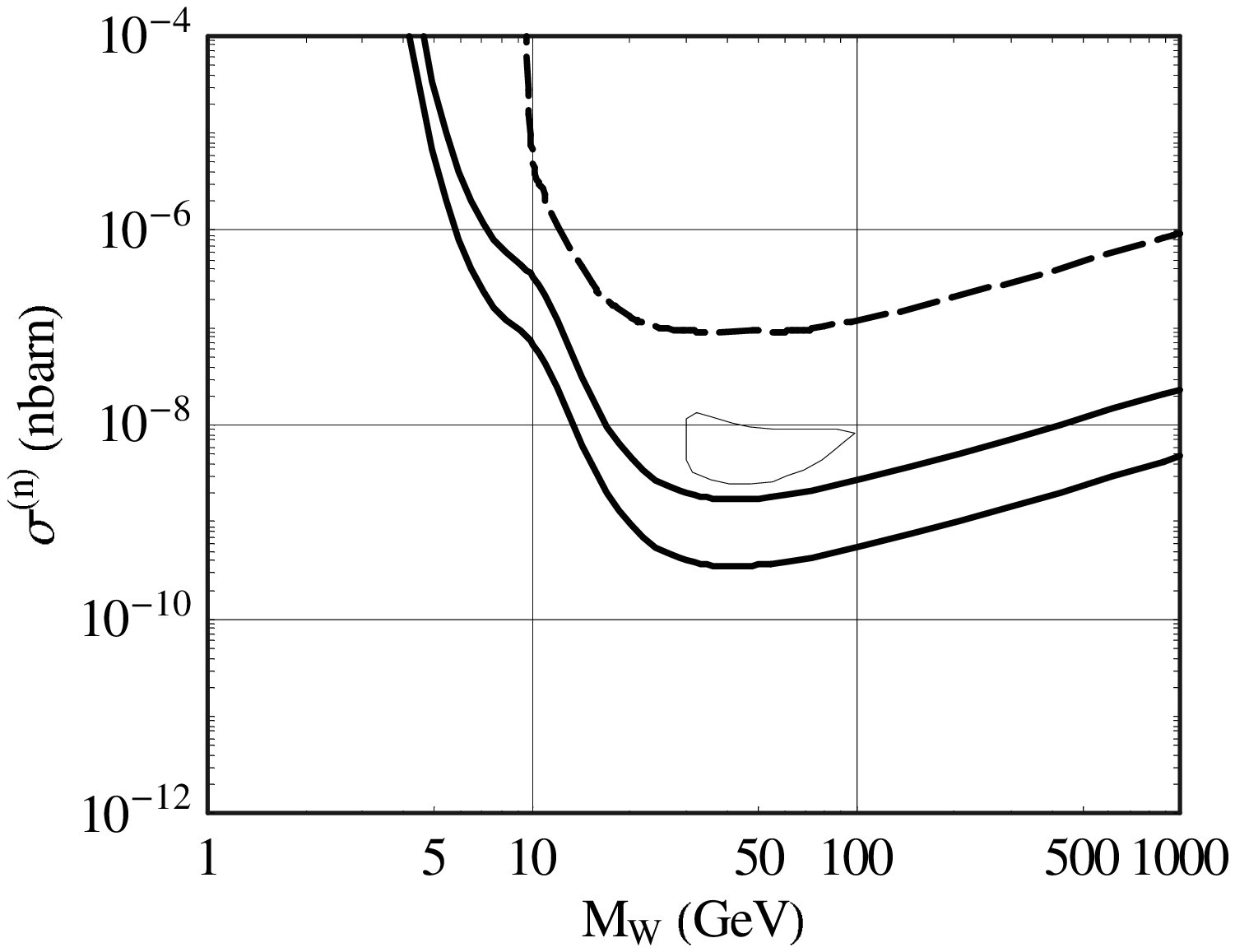}
\caption{Exclusion projected for 1 year of CUORE
assuming a threshold of 10 keV, a low energy resolution of 1 keV, and low energy background levels of 0.05 and 0.01 c/keV/kg/day
respectively. The closed curve represents the DAMA region. The dashed line corresponds to the current MiDBD result.}
\label{cuore_exclusion}
\end{center}
\end{figure}

In figure \ref{cuoricino_exclusion}, the exclusion plots for
coherent spin-independent WIMP-matter interaction are shown for
two possible values of the background of CUORICINO, 1 and 0.1
c/keV/kg/day. The first value is of the order of the background
already achieved from threshold onwards (10-50 keV) in the MiDBD
latest results (see \cite{Giulani}). The value 0.1 c/keV/kg/day is
a one-order-of-magnitude extrapolation from that currently
achieved in MiDBD (see discussion on Section 2) and is close to
the one obtained above 50 keV. In the case of CUORE, background
values of 0.05 and 0.01 c/keV/kg/day will be assumed. Notice,
moreover, that values of a few 0.01 c/keV/kg/day have been
obtained above 10 keV in the raw spectra of Germanium experiments
(like IGEX \cite{IGEXDM}) without using mechanisms of background
rejection, and so it does not seem impossible to achieve such
equivalent small values in crystal thermal detectors of tellurium
(only phonons, and no discrimination mechanism). To draw the two
exclusion contours of Fig. \ref{cuoricino_exclusion}, a low energy
resolution of 1 keV and an energy threshold of 10 keV have been
assumed as well as an exposure of 2 years of CUORICINO (81
kg$\cdot$year). The projected exclusion contours are compared with
the one currently obtained from MiDBD (dashed line). In figure
\ref{cuore_exclusion}, the exclusions for the two quoted values of
the background of CUORE, 0.05 and 0.01 c/keV/kg/day, are similarly
presented for an exposure of 1 year (760 kg$\cdot$year).

As previously noted, CUORE and to some extent CUORICINO have
detector masses large enough to search for the annual modulation
signal. As it is well known, an essential requirement to estimate
the prospects of any detector to search for annual modulation is
to have a superb control of systematic errors and to assure that
the stability of the various experimental parameters, which might
mimic periodic variations of the signals, are kept within a small
fraction of the (already tiny) expected signal. The various
changes of the set-up, crystals and shielding of the MiDBD
experiment have not provided a definitive estimation of the
long-term stability parameters of MiDBD. Possible instabilities
are that of the electronic gain and the ensuing time fluctuation
of the energy scale (both in energy thresholds and energy
resolutions), the temperature variations, the possible fluctuation
in time of the efficiency with which the triggered noise is
rejected and others. They must be kept well below the small
expected seasonal modulation of the WIMP signal. The fact that we
are dealing with a very small signal depending on time, which
typically amounts to a fraction between 1\% and 7\% of the average
count rates, reinforces the need for a control of the stability of
the experiment well below that range over long periods of time.
When assuming that all these fluctuations are controlled well
below the levels needed ($<$1\%), then one can proceed to analyze
the sensitivity of CUORICINO/CUORE to the annual modulation signal
on purely statistical grounds. This has been first attempted in
\cite{Ramachers} and \cite{Hasenbalg}, but a more extensive and
rigorous approach is followed in ref. \cite{Cebrian:2001qk} where
sensitivity plots for several types of detectors (and experimental
parameters) are presented, and in particular, for CUORE and
CUORICINO.

The sensitivity of a given experimental device to the annual
modulation signal (according to the detector material employed and
the experimental parameters of the detectors) has been extensively
studied in Ref. \cite{Cebrian:2001qk} on purely statistical
grounds.
Following the guidelines of that reference, it can be precisely
quantified by means of the $\delta$ parameter, defined from the
likelihood function or, equivalently, from the $\chi^2$ function
of the cosine projections of the data (for further details see
ref. \cite{Cebrian:2001qk}):

\begin{equation}\label{eq:asimptotic}
  \delta^2=y(\sigma=0)-y_{min}\simeq\chi^2(\sigma=0)-\chi^2_{min}.
\end{equation}

This parameter measures the statistical significance of the
modulation signal detected in an experimental set of data.
However, for a given ($m,\sigma$) and a given experiment the
expected value $\langle \delta^2 \rangle$ can be estimated using
the expression derived in ref. \cite{Cebrian:2001qk}:

\begin{equation}\label{expansion1}
  \langle \delta^2 \rangle=\frac{1}{2}\sum_{k}
  \frac{S_{m,k}(\sigma,m_W)^2\Delta
  E_k}{b_{k}+S_{0,k}}MT\alpha+2\label{eq:magic}.
\end{equation}

\noindent where $S_{m,k}$ and $S_{0,k}$ are the modulated and
non-modulated parts of the WIMP signal in the $k$th energy bin of
$\Delta E_k$ width, $b_k$ is the background in that energy bin and
$MT\alpha$ the effective exposure, being $\alpha$ a coefficient
accounting for the temporal distribution of the exposure time
around modulation maxima and minima ($\alpha = 1/n \sum_{i=1}^n
\cos^2 \omega (t_i -t_0)$ for $n$ temporal bins).

\begin{figure}[h]
\begin{center}
\includegraphics[width=1\textwidth]{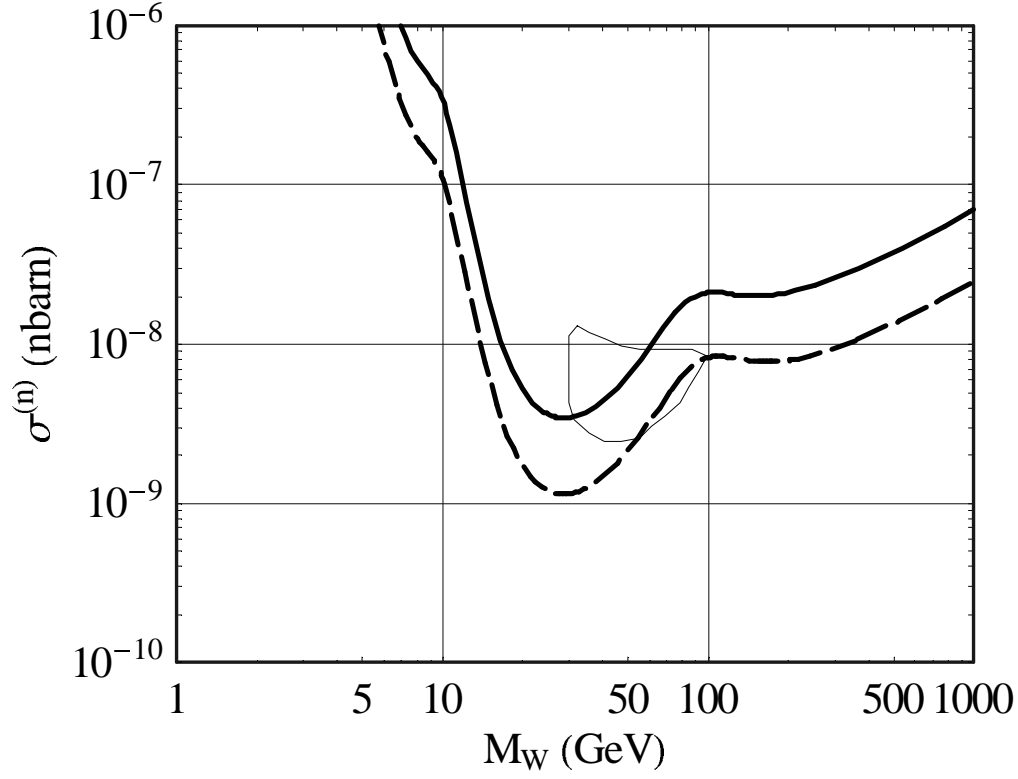}
\caption{Sensitivity plot in the ($m,\sigma$) plane for CUORICINO,
assuming a threshold of 10 keV, flat background $b=$ 1 (solid
line) and 0.1 c/keV/kg/day (dashed line) and two years of exposure
(81 kg year). It has been calculated for $\langle\delta^2\rangle=$
5.6 (see the text). The closed contour represents the 3$\sigma$ CL
region singled out by the modulation analysis performed by the
DAMA experiment \cite{Bernabei}.} 
\label{cuoricino_mod}
\end{center}
\end{figure}

\begin{figure}[h]
\begin{center}
\includegraphics[width=1\textwidth]{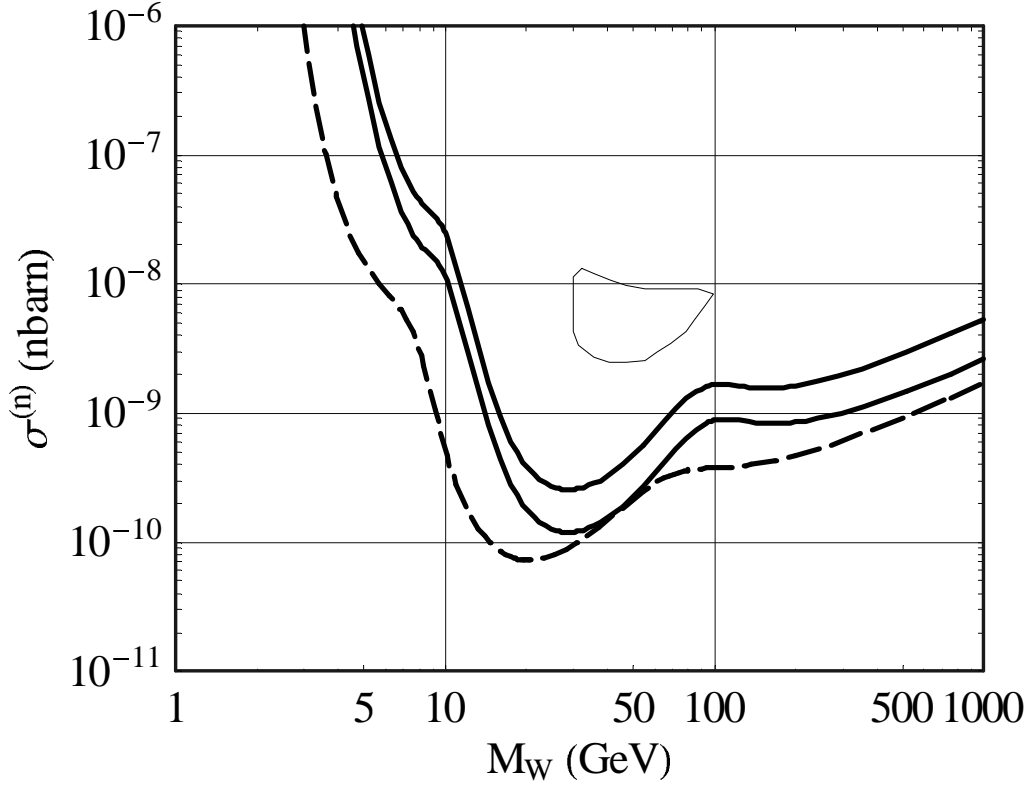}
\caption{The solid lines represent the sensitivity plot in the
($m,\sigma$) plane for CUORE, assuming a threshold of 10 keV, two
years of exposure (1500 kg year) and flat backgrounds of 0.05 and
0.01 c/keV/kg/day. It has been calculated for
$\langle\delta^2\rangle=$ 5.6 (see the text). The sensitivity
curve has been also calculated for a possible threshold of 5 keV
with a background of 0.01 c/keV/kg/day (dashed line). The closed
contour represents the 3$\sigma$ CL region singled out by the
modulation analysis performed by the DAMA experiment
\cite{Bernabei}.}
\label{cuore_mod}
\end{center}
\end{figure}

Using this equation we have estimated the region that could be
within reach for CUORE and CUORICINO with the above mentioned
assumptions on the background levels. We have fixed a value of 5.6
for $\langle \delta^2 \rangle$ that corresponds to 50\%
probability of obtaining a positive result at 90\% C.L.. In figure
\ref{cuoricino_mod} curves are shown obtained for a threshold of
10 keV, two years of exposure with CUORICINO (81 kg year) and two
assumed flat backgrounds of 1 and 0.1 c/keV/kg/day. One can see
that CUORICINO could already explore most of the DAMA region
looking for a positive annual modulation signal. In figure
\ref{cuore_mod} similar curves are presented, assuming flat
backgrounds of 0.05 and 0.01 c/keV/kg/day, two years of exposure
of CUORE (1500 kg year) and a threshold of 10 keV (solid lines).
The possibility of a lower thresholds of 5 keV with a background
of 0.01 c/keV/kg/day is also shown (dashed line).

In conclusion, CUORE and CUORICINO will be able to explore and/or
exclude WIMPs lying in large regions of their parameter space. The
capability of CUORICINO / CUORE to investigate the DAMA region
through the exclusion plot (time integrated method) relies in
getting a background of 0.1 c/keV/kg/day from 10 keV onwards,
independently of more elaborated time modulation methods which
require an exhaustive control of the stability of the experiment.
However, CUORICINO and CUORE could also attempt to look for annual
modulation of WIMP signals provided that the stability of the
experiment is sufficient.

\section{Solar axion detection} \label{axions}

Axions are light pseudoscalar particles which arise in theories in
which the Peccei-Quinn U(1) symmetry has been introduced to solve
the strong CP problem \cite{Peccei:1977hh}. They could have been
produced in early stages of the Universe being attractive
candidates for the cold dark matter (and in some particular
scenarios for the hot dark matter) responsible to 1/3 of the
ingredients of a flat universe. Dark matter axions can exist in
the mass window $10^{-2(3)}$ eV $<m_{a}\leq 10^{-6}$ eV, but
hadronic axions could exist with masses around the eV.

Axions could also be copiously produced in the core of the stars
by means of the Primakoff conversion of the plasma photons. In
particular, a nearby and powerful source of stellar axions would
be the Sun. The solar axion flux can be easily estimated
\cite{vanBibber:1989ge,Creswick} within the standard solar model,
resulting in an axion flux of an average energy of about 4 keV
that can produce detectable X-rays when reconverted again in an
electromagnetic field. Moreover, it has been pointed out recently
that the dimming of supernovae SNIa might be due to the conversion
of photons into axions in the extra-galactic magnetic field
\cite{Csaki}. A photon-axion ($\gamma$-a) oscillation could make
unobservable about 1/3 of the SN emitted light and so, they would
appear fainter than implied by the luminosity-distance versus
redshift relation, without need to invoke an accelerated expansion
of the Universe. The SN result would be matched by axions of mass
$\sim10^{-16}$ eV and coupling to photons $\gagamma \sim2.5 \times
10^{-12}$ GeV$^{-1}$. So stellar axions may play an important role
in Cosmology. We would like to stress that, although we focus on
the axion because its special theoretical motivations, all this
scenario is also valid for any generic pseudoscalar (or scalar)
particle coupled to photons \cite{Masso:1995tw}. Needless to say
that the discovery of any type of pseudoscalar or scalar particle
would be extremely interesting in Particle Physics. We will keep
our discussion, however, restricted to the case of solar axions.

Crystal detectors provide a simple mechanism for solar axion
detection \cite{Paschos,Creswick}. Axions can pass in the
proximity of the atomic nuclei of the crystal where the intense
electric field can trigger their conversion into photons. The
detection rate is enhanced if axions from the Sun coherently
convert into photons when their incident angle with a given
crystalline plane fulfills the Bragg condition. This induces a
correlation of the signal with the position of the Sun which can
be searched for in the data and allows for background subtraction.
The potentiality of Primakoff conversion in crystals relies in the
fact that it can explore a range of axion masses ($m_a\gsim 0.1$
keV) not accessible to other direct searches. Moreover it is a
relatively simple technique that can be directly applied to
detectors searching for WIMPs.

Primakoff conversion using a crystal lattice has already been
employed in two germanium experiments: SOLAX \cite{SOLAX} and
COSME-II \cite{COSME} with the ensuing limits for axion-photon
coupling $\gagamma \lsim 2.7\times 10^{-9}$ GeV$^{-1}$ and
$\gagamma \lsim 2.8\times 10^{-9}$ GeV$^{-1}$ respectively. Also
the DAMA collaboration has analyzed 53437 kg-day of data of their
NaI set up \cite{dama_axion}, in a search for solar axions,
following the techniques developed in ref. \cite{Cebrian}, where a
calculation of the perspectives of various crystals detectors
(including NaI) for solar axion searches has been made. The DAMA
result $\gagamma \lsim 1.7\times 10^{-9}$ GeV$^{-1}$ improves
slightly the limits obtained with other crystal detectors
\cite{SOLAX,COSME} and agrees with the result predicted in ref.
\cite{Cebrian}. These "crystal helioscopes" constraints are
stronger than that of the Tokyo axion helioscope \cite{tokyo} for
$m_a \gsim 0.26$ eV and do not rely on astrophysical
considerations (i.e. on Red Giants or HB stars dynamics
\cite{Raffelt}). The orientation of the crystal was not known so
that the data were analyzed taking the angle corresponding to the
most conservative limit.

It has been noted that the model that yields the solar axion
fluxes used to calculate the expected signals is not compatible
with the constraints coming from helioseismology if $\gagamma
\gsim 10^{-9}$ GeV$^{-1}$ \cite{Schlattl}. This would imply a
possible inconsistency for solar axion limits above that value,
and sets a minimal goal for the sensitivity of future experiments.

The use of CUORE to search for solar axions via Bragg scattering
should have a priori some advantages with respect to germanium
detectors, because of the larger mass and the known orientation of
the crystals. On the other hand, as the cross-section for
Primakoff conversion depends on the square of the atomic number,
TeO$_{2}$ will be a priori a better candidate than Germanium.
Needless to say that a low energy threshold is mandatory because
the expected signal lies in the energy region 2 keV $\lsim E
\lsim$ 10 keV and is peaked at $E\simeq 4$ keV.

A detailed analysis has been performed \cite{Cebrian} for a
TeO$_2$ crystal (which has a tetragonal structure \cite{Crystal})
assuming different values for the experimental parameters. As it
is shown in Ref. \cite{Cebrian}, the bound on axion-photon
coupling which a given experiment can achieve can be estimated
through the expression:

$$ g_{a\gamma \gamma }<g_{a\gamma \gamma }^{\lim }\simeq k\left(
\frac b{\rm c/keV/kg/day}\frac{\rm kg}M\frac{\rm years}T\right)
^{1/8}\times 10^{-9}{\rm \ GeV}^{-1} $$

\noindent where $k$ depends on the crystal structure and material,
as well as on the experimental threshold and resolution. For the
case of TeO$_2$ and a threshold of 5 keV, $k$ has been calculated
to be $k=2.9$ assuming an energy resolution of 1 keV. The
computation of this expression for some assumed values of the
experimental parameters is shown in table \ref{tab_cuore} for
CUORICINO and CUORE. In all cases flat backgrounds and 2 years of
exposure are assumed.

\begin{figure}[h]
\begin{center}
\includegraphics[width=0.95\textwidth]{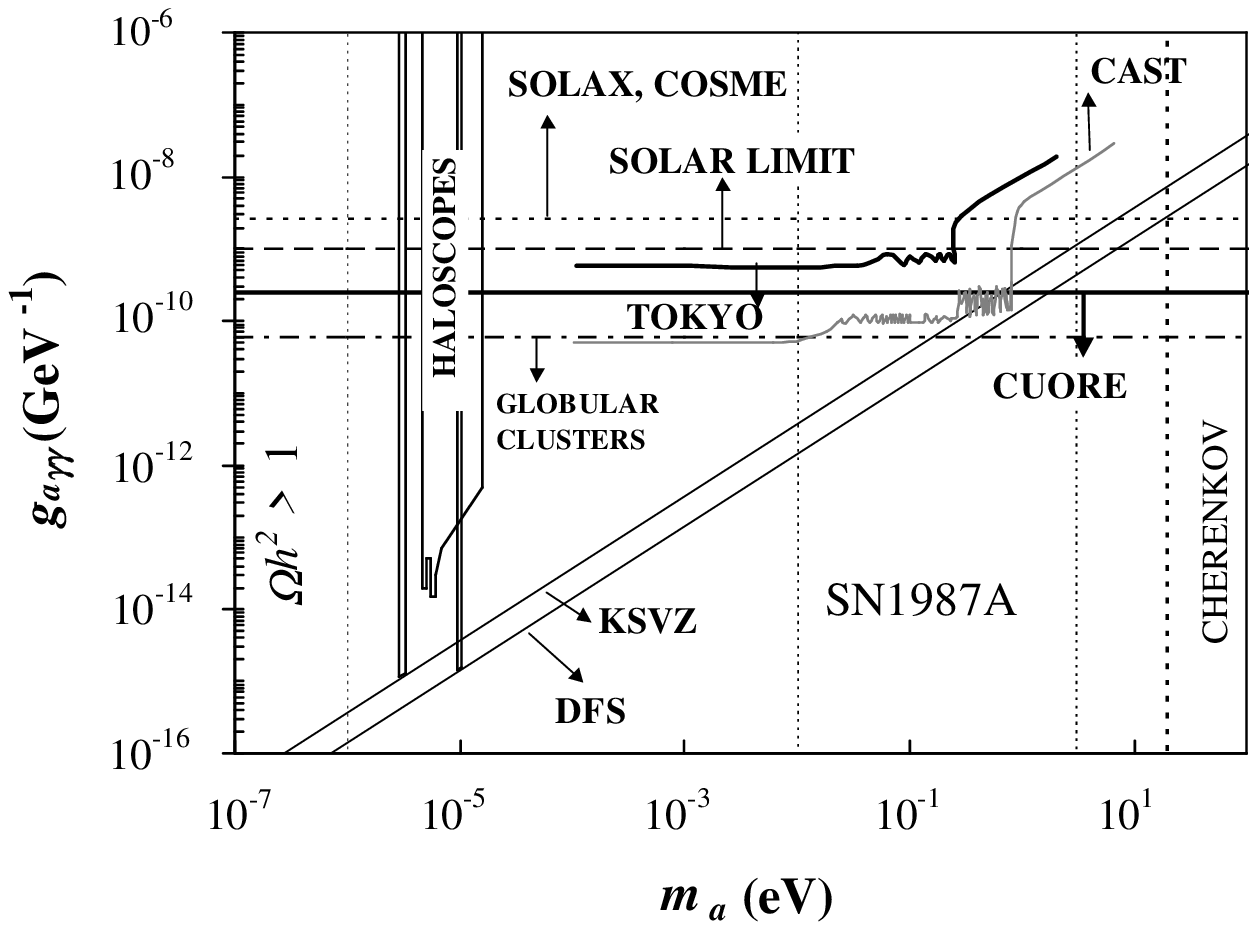}
\caption{Best bound attainable with CUORE (straight line labelled
"CUORE") compared with others limits.}
\label{fig}
\end{center}
\end{figure}

It is worth noticing the faible dependence of the ultimate
achievable axion-photon coupling bound on the experimental
parameters, background and exposure MT: the 1/8 power dependence
of $g_{a\gamma\gamma}$ on such parameters softens their impact in
the final result. The best limit shown in table \ref{tab_cuore} is
in fact only one order of magnitude better than the present limits
of SOLAX and COSME-II. The $\gagamma$ bound that CUORE could
provide is depicted comparatively to other limits in figure
\ref{fig}.

The limit which can be expected from the CUORICINO experiment is
comparable to the helioseismological bound mentioned before (see
Table \ref{tab_cuore}). CUORE could go even further (see Figure
\ref{fig} and Table \ref{tab_cuore}). Notice that in both cases an
energy threshold of $E_{thr}\sim5$ keV (and resolution of $\sim$1
keV) has been assumed. That value will be confirmed only after
knowing the performances and preliminary runs of CUORICINO. As was
described at length in Ref. \cite{Cebrian}, the crucial parameters
for estimating the perspectives on solar axion detectors with
crystals rely on the energy threshold and resolution (appearing in
$k$), and the level of background achieved (although the influence
of this parameter is damped by a factor 1/8). In particular, a
threshold of 5-8 keV would loose most of the axion signal. Other
crystal detectors with, say, Ge or NaI (GENIUS, MAJORANA, GEDEON,
DAMA, LIBRA, ANAIS, \dots) could surpass CUORE as axion detectors
because the energy thresholds of these projects are supposed to be
significantly lower. Also the background is expected to be better.
Nevertheless, it should be stressed that the bounds on $g_{a\gamma
\gamma}$ obtained with this technique in the various proposed
crystal detector arrays stagnate at a few $\times 10^{-10}$
GeV$^{-1}$, not too far from the goal expected for CUORE, as has
been demonstrated in \cite{Cebrian}. There are no realistic
chances to challenge the limit inferred from HB stars counting in
globular clusters \cite{Raffelt} and a discovery of the axion by
CUORE would presumably imply either a systematic effect in the
stellar-count observations in globular clusters or a substantial
change in the theoretical models that describe the late-stage
evolution of low-metallicity stars. To obtain lower values of
$\gagamma$ one should go to the magnet helioscopes like that of
Tokio \cite{tokyo} and that of CERN (CAST experiment \cite{cast}
currently being mounted). In particular, the best current
experimental bound of $\gagamma$ published comes from the Tokyo
helioscope: $\gagamma \leq 6 \times 10^{-10}$ GeV$^{-1}$ for
$m_{a} \lesssim 0.03$ eV and $\gagamma \leq 6.8-10.9 \times
10^{-10}$ GeV$^{-1}$ for $m_{a} \sim 0.05-0.27$ eV. The
sensitivity of CAST is supposed to provide a bound $\gagamma \leq
5 \times 10^{-11}$ GeV$^{-1}$ or even lower. A recent, preliminary
run of CAST, with only $\sim$5 hours of "axion light" has improved
already the Tokyo limit.

\section{Conclusions}

We have reported the perspectives of CUORE, a projected massive
760 kg array of 1000 TeO$_2$ bolometers, and of its first stage
CUORICINO, with 40 kg of the same crystals, as far as their physics potential to detect various
types of rare events is concerned. The estimated background and
resolution, based on Monte Carlo studies, together with the
results obtained in recent improvements in the performances of the
MiDBD experiment and the information obtained from the
preliminary tests of CUORICINO, have allowed us to assess the
potentialities of these experiments for double beta decay
searches, solar axion detection and WIMP exclusion or
identification. In these three types of searches, CUORE and to
some extent CUORICINO will be powerful tools to explore, with
higher sensitivity, such rare phenomena.

\section{Acknowledgements}

This work has been partially supported by the Spanish CICYT
(contract AEN99-1033), Italian INFN, the US National Science
Foundation and the EU Network (Contract ERB-FMRX-CT-98-0167).
Thanks are due to our student F. Capozzi for the work done on the
MC simulations, and to G. Luzon for discussions on the cosmogenic
activation of the crystals.





\newpage

\begin{table}[h]
\centering
  \begin{tabular}{cccc} \hline\hline
    $^{76}$Ge  &  $^{130}$Te &  $^{136}$Xe & nuclear model  \\ \hline
     $1.12\times 10^{-13}$ &  $5.33\times 10^{-13}$ & $1.18\times 10^{-13}$ &  QRPA \cite{QRPA1} \\
     $1.12\times 10^{-13}$ &  $4.84\times 10^{-13}$ & $1.87\times 10^{-13}$ &  QRPA \cite{QRPA2} \\
     $1.87\times 10^{-14}$ &  $3.96\times 10^{-13}$ & $7.9\times 10^{-14}$ &  QRPA \cite{QRPA3} \\
     $1.54\times 10^{-13}$ &  $1.63\times 10^{-12}$ &                        &  Weak Coupling SM \cite{WCSM} \\
     $1.13\times 10^{-13}$ &  $1.1\times 10^{-12}$ &                         &  Generalized Seniority \cite{GenSen} \\
     $1.21\times 10^{-13}$ &  $5.0\times 10^{-13}$ & $1.73\times 10^{-13}$ &  QRPA \cite{QRPA4} \\
     $7.33\times 10^{-14}$ &  $3.0\times 10^{-13}$ & $1.45\times 10^{-13}$ &  QRPA without pn pairing \cite{QRPA5} \\
     $1.42\times 10^{-14}$ &  $1.24\times 10^{-13}$ & $9.3\times 10^{-14}$ &  QRPA with pn pairing \cite{QRPA5} \\
     $5.8\times 10^{-13}$ &  $3.18\times 10^{-12}$ &                        &  \cite{Kla} \\
     $1.5\times 10^{-14}$ &  $4.52\times 10^{-14}$  & $2.16\times 10^{-14}$ &  Large basis SM \cite{LbSM} \\
     $9.5\times 10^{-14}$ &  $3.6\times 10^{-13}$ & $6.06\times 10^{-14}$ &  Operator expansion method \cite{OEM} \\
     \hline\hline
  \end{tabular}
  \caption{$2\beta 0 \nu$ nuclear merits $F_N^{0\nu}$ (y$^{-1}$)
  of emitters used in some source=detector calorimeters, according
  to various nuclear models.}\label{tab_doblebeta}
\end{table}

\newpage

\begin{table}[h] \centering
  \begin{tabular}{ccccc} \hline\hline
    {\bf Mass}  &  {\bf Resolution}&  {\bf Threshold}   & {\bf Background } &  {\bf $g_{a\gamma \gamma }^{\lim }$ (2 years)}\\
     {\bf (kg)}  &  {\bf (keV)}  &  {\bf (keV)} & {\bf (c/kg/keV/day)} &  {\bf (GeV$^{-1}$)}\\ \hline
40.7       &  1   & 5    &  0.1  &  1.3$\times 10^{-9}$\\

40.7 & 1 & 5    &  1  &  1.7$\times 10^{-9}$\\
     750       &  1   & 5    &  0.01  &  6.5$\times 10^{-10}$\\
     750       &  1   & 5    &  0.05  &  8.0$\times 10^{-10}$\\
     \hline\hline
  \end{tabular}
    \caption{Expected limits on the photon-axion coupling for 2 years of exposure of CUORICINO and CUORE assuming the quoted values
  for the experimental parameters}\label{tab_cuore}
\end{table}





\newpage

\newpage

\newpage

\newpage

\newpage

\end{document}